\documentclass[%
article,letterpaper,
superscriptaddress,
 amsmath,amssymb,
 aps,
]{revtex4}
\usepackage{tocvsec2}
\usepackage[utf8]{inputenc}
\usepackage{hyperref}
\hypersetup{
    colorlinks=true,
    linkcolor=blue,
    filecolor=magenta,      
    urlcolor=magenta,
    citecolor=blue
}
\usepackage[top=1in, bottom=1.25in, left=1.25in, right=1.25in]{geometry}
\usepackage{titlesec}
\usepackage{enumerate}
\usepackage{enumitem}
\usepackage{bm}
\usepackage{amsfonts}
\usepackage{amsmath}
\usepackage{amssymb}
\usepackage{etoolbox}

\titlespacing*{\chapter}{0pt}{-50pt}{20pt}
\titleformat{\chapter}[display]{\normalfont\large\bfseries}{\chaptertitlename\ \thechapter}{20pt}{\Large}
\titleformat*{\section}{\fontsize{16}{20}\selectfont}

\def\ltap{\ \raise.3ex\hbox{$<$\kern-.75em\lower1ex\hbox{$\sim$}}\ }

\newcommand{\nub}{\overline{\nu}}

\newcommand{\FNAL}{Fermi National Accelerator Laboratory, Batavia, 
Illinois 60510, USA}

\begin{document}

\title{Summary of the NuSTEC Workshop on Neutrino-Nucleus Pion Production in the Resonance Region}


\author{L.~Aliaga} \affiliation{\FNAL}
\author{A.~Ashkenazi} \affiliation{Massachusetts Institute of Technology, Cambridge, Massachusetts 02139, USA}
\author{C.~Bronner} \affiliation{Kamioka Observatory, Institute for Cosmic Ray Research, University of Tokyo, Kamioka, Gifu, Japan}
\author{J.~Calcutt} \affiliation{Michigan State University, East Lansing, MI 48824, USA}
\author{D.~Cherdack} \affiliation{University of Houston, Houston, TX 77204, USA}
\author{K.~Duffy} \affiliation{\FNAL}
\author{S.~Dytman} \affiliation{University of Pittsburgh, Pittsburgh, PA, 15260, USA}
\author{N.~Jachowicz} \affiliation{Department of Physics and Astronomy, Ghent University, B-9000 Gent, Belgium}
\author{M.~Kabirnezhad} \affiliation{University of Oxford, Oxford OX1 3RH, United Kingdom}
\author{K.~Kuzmin} \affiliation{Joint Institute for Nuclear Research, Dubna, Russia}
\author{G.~A.~Miller} \affiliation{University of Washington, Seattle, WA 98195,USA  }
\author{T.~Le} \affiliation{Physics Department, Tufts University, Medford, Massachusetts 02155, USA}
\author{J.~G.~Morf\'{i}n} \affiliation{\FNAL}
\author{U.~Mosel} \affiliation{Institut für Theoretische Physik, Universität Giessen, D-35392 Giessen, Germany}
\author{J.~Nieves} \affiliation{Instituto 
de F\'\i sica Corpuscular (centro mixto CSIC-UV),
Institutos de Investigaci\'on de Paterna, Aptdo. 22085, 46071, Valencia, Spain}
\author{K.~Niewczas} \affiliation{Department of Physics and Astronomy, Ghent University, B-9000 Gent, Belgium} \affiliation{Institute of Theoretical Physics, University of Wroc{\l}aw, Plac Maxa Borna 9, 50-204, Wroc{\l}aw, Poland}
\author{A.~Nikolakopoulos} \affiliation{Department of Physics and Astronomy, Ghent University, B-9000 Gent, Belgium}
\author{J.~Nowak} \affiliation{Department of Physics, Lancaster University, LA1 4YB, Lancaster, United Kingdom}
\author{J.~Paley} \affiliation{\FNAL}
\author{G.~Pawloski} \affiliation{School of Physics and Astronomy, University of Minnesota Twin Cities, Minneapolis, Minnesota 55455, USA}
\author{T.~Sato} \affiliation{Research Center for Nuclear Physics, Osaka University, Ibaraki, Osaka, 567-0047, Japan}
\author{L.~Weinstein} \affiliation{Old Dominion University, Norfolk, Virginia 23529, USA}
\author{C.~Wret} \affiliation{University of Rochester, Rochester, New York, U.S.A.}

\date{\today}

\begin{abstract}
The   \href{https://indico.fnal.gov/event/20793/overview}{NuSTEC workshop} held at the University of Pittsburgh in October 2019 brought theorists and experimentalists together to discuss the state of modeling and measurements related to pion production in neutrino-nucleus scattering in the kinematic region where 
pions are produced through both resonant and non-resonant mechanisms. Modeling of this region is of critical importance to the current and future accelerator- and atmospheric-based neutrino oscillation experiments.  For the benefit of the community, links to the presentations are accompanied by annotations from the speakers highlighting significant points made during the presentations and resulting discussions.
\end{abstract}

\maketitle

\newpage

\tableofcontents

\newpage

\section[Executive Summary]{Executive Summary}
Current and future neutrino oscillation experiments rely on neutrino interaction simulations to relate observed kinematic and calorimetric variables to the neutrino energy.  The simulations are also used to correct for detector acceptance and background predictions, which depend on the kinematics of final-state particles.  The uncertainties related to the modeling of neutrino-nucleus interactions result in some of the larger systematic uncertainties in neutrino oscillation measurements.  Many oscillation measurements depend on a strong knowledge of interactions where nucleon resonance production critically influences the composition of the final state. 



The T2K and H2K experiments use water Cherenkov detectors to observe neutrino oscillations, and can only observe the kinematics of particles with energies above the Cherenkov light threshold.  T2K uses primarily carbon-based measurements in its near detector to constrain the rate and kinematics of particles produced with energies below the Cherenkov threshold.  Modeling the differences between $\Delta$ resonance production in neutrino-carbon and neutrino-oxygen interactions is therefore important for the T2K and H2K measurements.

The NOvA and DUNE experiments use neutrino beam energies peaked 
somewhat above the $\Delta$ resonance 
energies and are thus also sensitive to both higher-mass resonant and non-resonant pion production. This kinematic region is currently poorly studied experimentally and the few available models are essentially untested. These experiments use liquid scintillator (LS, NOvA) and liquid argon (LAr, DUNE) tracking calorimeters to relate the incoming neutrino energy to the kinematic reconstruction of particles and calorimetric energy depositions.  Although the near and far detectors used in these oscillation experiments have the same nuclear target, the 
acceptance and incoming neutrino flux energy distribution
is quite different between the two detectors.  Modeling the rates and kinematics of the final-state particles produced is therefore important for the NOvA and DUNE measurements.

The Fermilab Short-baseline Neutrino (SBN) Program utilizes LAr TPCs exposed to a beam of neutrinos with a flux similar to that of T2K/H2K.  Although the SBN experiments plan to measure cross sections, these measurements will be at the right energies but different nucleus for T2K/H2K, and lower energies but same nucleus for DUNE.  


A number of issues were evident before the workshop:  
\begin{itemize}
    \item What is the best way to blend resonant and nonresonant pion production?
    \item What can be learned from the pion scattering and pion electro-production experiments and interpretations in the past?
    \item What is the best way to mix low-energy resonant processes with high-energy deep inelastic (DIS) processes?
    \item What are possible contributions from lattice QCD?
\end{itemize}


Theorists and experimentalists with experience working on topics related to lepton-nucleus 
pion production gathered at the the \href{https://indico.fnal.gov/event/20793/overview}{NuSTEC workshop} held at the University of Pittsburgh in October 2019.  During the 4-day long workshop, presentations at the workshop covered the impact of pion production modeling and uncertainties in neutrino experiments, existing models and new model development, implementation of models in neutrino event generators, and relevant neutrino, electron and pion scattering measurements.  Several hours per day were dedicated to in-depth questioning and discussions. These presentations and discussions were important to define future work to address the problems listed above.

Some of the discussion at the workshop centered around clarification of statements made by the presenters, but most discussions centered around the various challenges of modeling pion production in the full resonance region including the transition to DIS pion production, and the challenges associated with proposals for future measurements of pion production in active collaborations analyzing data.  There was clear consensus that resonance production measurements for both electron and neutrino beams are very important for the future success of oscillation experiments.  
Discussions at the workshop also concluded that theory benefits greatly from past electroproduction analyses, but must also be expanded.  It was also made evident that the understanding of electroproduction data is not sufficient for the present needs.

NuSTEC has the goal of organizing workshops and schools to advance knowledge of neutrino interactions and related subjects. The kinematic region covered by this workshop on pion production in the resonance region blends quite seamlessly with the prior year \href{https://indico.cern.gov/}{NuSTEC workshop on Shallow and Deep Inelastic Scattering} \cite{ref:NuSTEC2018}.  Taken together these workshops emphasize that although the $\Delta P_{33}(1232)$ resonance is the focus of many studies, the kinematic region with mass above the $\Delta$ is in great need of both experimental and theoretical study.  This region is currently an unknown mixture of higher W resonances, non-resonant pion production and the start of deep-inelastic scattering.  Even attempts to understand the inclusive behavior in this region using quark-hadron duality and results from the better-studied DIS region have not been successful.  

That over 50\% of the expected DUNE events are from this poorly studied higher-W regime  emphasizes the real need to devote increased effort to explore this region.  Having brought together the worlds' foremost researchers in theory, experiment and simulation from both the high-energy and nuclear communities, this workshop fostered in depth discussions that helped bridge the communication gaps between these communities opening the way to this necessary exploration.


We are very grateful to the Pittsburgh Particle Physics Astrophysics and Cosmology Center (PittPAC) for their financial and administrative support of the workshop.  This support enabled many speakers and early-career scientists to attend the workshop, creating a dynamic and inclusive experience for all.



\newpage


\section{Introduction and Overview}
\subsection[Introduction to RES and Non-RES Theory and Models - Steve Dytman]{Introduction to RES and Non-RES Theory and Models - Steve Dytman} \label{sec:1a}
\begin{center}
[\href{https://indico.fnal.gov/event/20793/contribution/2/material/slides/0.pdf}{Presentation}]   
\end{center}

This talk gave an overview of resonance production from electron and neutrino beams.  Each has a particular history, each interesting in their own way.  Electron beams are more intense and are monoenergetic.  Although all existing neutrino experiments suffer from a lack of statistics, they have the advantage of using general purpose detectors with a very large acceptance.

For nucleon targets, the electron data is of very high quality and $(e,e'\pi)$ experiments have been interpreted for resonance content by the MAID group and others.  The DCC results are presented at this workshop by Toru Sato.  The only neutrino data comes from bubble chamber data from at least 30 years ago.  They have low statistics and broad energy range, which makes the data hard to interpret. Various theoretical models are available and were presented at the workshop.  The DCC model has the advantage of giving good agreement with a wide range of both electron and neutrino data.  Generators tend to use the Rein-Sehgal or Berger-Sehgal models (presented by Konstantin Kuzmin) which are simpler to implement.

For nuclear targets, data is sparse for both probes.  Recent neutrino data from MiniBooNE, MINERvA, Argoneut, and T2K are of low statistics and workers are still exploring the consistency among the data sets.  A key missing data set is $(e,e'\pi)$ in nuclei.  The $e4\nu$ experiment (Larry Weinstein) is analyzing electron-nucleus data with pions in the final state.  Another interesting direction comes from the new experiments for  pion-nucleus interactions.

Theory models for nuclear targets use the nucleon target data with the impulse approximation. Nuclear models tend to be local Fermi gas (LFG) for both electron and neutrino experiments.  Nuclear effects such as $NN$ correlations and medium corrections haven't been commonly applied.  The transition between the nucleon-based and quark-based models remains a frontier.

\subsection[What We have Learned from e-A and pi-A Measurements - Gerald A. Miller]{What We have Learned from e-A and pi-A Measurements - Gerald A.  Miller} \label{sec:1b}
\begin{center}
[\href{https://indico.fnal.gov/event/20793/contribution/3/material/slides/0.pdf}{Presentation}]    
\end{center}
I was asked to talk about the history of pion-nucleus interactions with a focus on Delta excitation and propagation in the nucleus. So the talk presented a review of pion-nucleus phenomenology, a discussion of the meaning of a resonance and some general remarks. 

Pions are made  when the leptonic interaction with a nucleon produces a resonance, which decays by emitting pions and also by quasi-eleastic production in which a fast nucleon emits a pion. In general there are many coherent ways to make a final state  involving a pion,  and the different amplitudes should be added and the result squared to compute a cross section.

The talk relied on the references \cite{Ericson:1988gk} and \cite{Lee:2002eq}. Pion nucleon and pion nucleus total cross sections are dominated by the Delta resonant peak for pion kinetic energies between 50 and 400 MeV. The peak broadens and moves down in energy  as the value of $A$ increases. The total cross sections consist three important contributions: elastic, inelastic (mainly quasielastic knockout)  and absorption (pion disappears). All of these increase with $A$ but the absorption cross section rises faster than the total.

The $\Delta$ is made in the nucleus and moves through it. A natural question to ask is what is the mass and width of this particle in the nucleus. The $\Delta$ experiences a mean field potential. It's width increases due to pion absorption but also decreases due to Pauli blocking of the $N\pi$ decay channel. These effects tend to cancel. The $\Delta$ also experiences multiple scattering.  $\Delta$ propagation was an important topic in the 1970's and 1980's. The talk discussed Refs.~\cite{Kisslinger:1973np,Kisslinger:1975gc,Oset:1979cx,Oset:1979tk,Oset:1979bi,Hirata:1977is,Hirata:1978wp,Freedman:1981dz,Freedman:1982yp}. The results showed generally that $\Delta$'s were modified in the medium, with the attractive nature of the real part being well-established.  The work that described  nuclei form $^{16}$O to $^{208}$Pb was that of Refs.~\cite{Freedman:1981dz,Freedman:1982yp}. They parameterized the $\Delta$-nucleus interaction as $(V+i W)/T_R=a +b (T_\pi/T_R)+c(T_\pi/T_R)^2$, with $T_R=180$ MeV, $a=0.44-0.77i,\,b=-11.24+1.24 i,\,c=0.47-0.47i$. This to be used in the resonance region only. This was in qualitative agreement with Refs.~\cite{Hirata:1977is,Hirata:1978wp}.
An important caution is that the elastic scattering angular distributions can be well described using black disk scattering~\cite{Zeidman:1978qp}.

Pion absorption is an important process because with it the production of a resonance does not lead to pions in the final state. Absorption is often a multi-nucleon process because the elastic $\pi N$ angular distribution is backward peaked. When a pion scattering backward it loses energy, sometimes knocking a nucleon out of the nucleus. This could happen several times making it easier for the pion to be absorbed. The final states produced by pion absorption were studied with the LADS detector~\cite{Kotlinski:2000hp}.  For $\pi^+$ absorption the $pp$ final state accounts for only 23 \% of the total absorption, $2pn$ is 35 \%, $3p$ is 13 \% and $3pn$ is 9 \%. These data suggest that final state interactions are the dominant mechanism.

The summary of of pion-nucleus interactions is provided here. The total cross section is roughly equal parts: elastic, quasi-free knockout, and absorption. 
Absorption involves more than 2 nucleons  with about a 77 \% probability. The detailed mechanism is not established.
Nuclear cross sections do not generally vary as A$^{2/3}$. The nuclear physics of the $\Delta$ is important.

Some time was spent discussing  the true nature of the $\Delta$ as a resonance in pion-nucleon scattering~\cite{Theberge:1980ye,Thomas:1981vc}, in which the 3-quark $\Delta$ occupies only 80 \% of the physical state. The remainder is a pion-nucleon component. The $\Delta$ is the simplest resonance; all the others are more complicated. This complicates the nuclear interactions. The resonance structure affects pion production cross sections. Experimenters use data instead of theory to model the events. 
But there is a  worry:
Different reaction mechanisms can reach the same final state. In that case there is  quantum interference and then one 
must add amplitudes and then square. Details are needed to do that correctly. Once  a pion is made it can do all the things discussed previously.

The talk also contains some general remarks about potential ways to improve the modelling of events. In particular, it is necessary to take the granularity (nucleon-nucleon correlations)  of the nuclear density into account.

 I wish to make a final remark about a subject that was often discussed in the workshop: the axial form factor of the nucleon. We  have recently produced a paper~\cite{Zhang2019:xxx} on a  unified model of nucleon elastic form factors with  implications for neutrino-oscillation experiments.  We studied
vector and axial  form factors,
using the light-front approach to build a quark-diquark model of the nucleon
with an explicit pion cloud. The light-front wave functions were  calibrated to existing experimental
information on the nucleon's electromagnetic form factors, and then used to
predict the axial form factor. We predict the squared charge radius of the
axial pseudo-vector form factor to be $r_A^2 = 0.29 \pm 0.03\, \mathrm{fm}^2$,
where the small error accounts for the model's parametric uncertainty. We use
our form factor   to explore the (quasi-)elastic scattering of neutrinos
by (nuclei)nucleons, with the result that the widely-implemented dipole ansatz is found to be an
inadequate approximation. 
The dipole approximation leads to a 5-10 \% over-estimation of the total cross section,
depending on the (anti)neutrino energy. Over-estimations of similar
size in the flux-averaged cross sections for the upcoming DUNE long-baseline
neutrino-oscillation experiment are also predicted.


\section{Impact of Pion Production Modeling in Neutrino Experiments}

\subsection[Impact of RES modeling to the sub-GeV global neutrino oscillation program - Jaroslaw Nowak]{Impact of RES modeling to the sub-GeV global neutrino oscillation program - Jaroslaw Nowak} \label{sec:2a}
\begin{center}
[\href{https://indico.fnal.gov/event/20793/contribution/4/material/slides/0.pdf}{Presentation}]
\end{center}

The presentation describes the effects of the modelling of pion production channels and reinteraction in the nucleus on the sensitivities and results of experiments with the sub-GeV neutrino beams. 
Currently, there are two neutrino beams with mean neutrino energy below 1~GeV: Booster Neutrino Beamline (BNB) \cite{AguilarArevalo:2008yp} and T2K beam \cite{Abe:2012av}. The experiments on the BNB investigate the short-baseline oscillations. The T2K primary goal is the investigation of the long-baseline neutrino oscillations, but they also reported on the short-baseline appearance study using the near detector (ND280) \cite{Abe:2014nuo}.

In the BNB and T2K beam, the dominant neutrino interaction type is the quasi-elastic scattering. The pion production contributes up to about 25\% with a significant part of the interactions coming from the high energy tail of the beams.

In the sub-GeV pions are mainly produced via the resonance excitation mechanism. There is also a few per cent contribution from non-resonant background \cite{Fogli:1979}. The MiniBooNE \cite{AguilarArevalo:2009eb, AguilarArevalo:2010bm, AguilarArevalo:2010xt} and T2K \cite{Abe:2016tmq, Abe:2016aoo, Abe:2017rfw, Abe:2019arf} published results for the pion production channels that are very useful for the oscillation experiments as they allow for better handling of systematic uncertainties. Also, the MINER$\nu$A experiment published results for pion production channels \cite{McGivern:2016bwh, Altinok:2017xua, Le:2019jfy}. Although the results from the three experiment are useful, they also introduced some confusions as there are tensions between the results.

The first ground-breaking experiment in the neutrino interactions physics in the sub-GeV region was MiniBooNE. This was mainly due to the high statistic of data they collected \cite{AguilarArevalo:2007it} and the topology-based measurements they adopted. That was one of the attempts to avoid model dependency in the neutrino interactions measurements. 

The T2K and BNB experiments have used four of the Monte Carlo generators: NUANCE, GENIE,  NEUT and NuWro. NUANCE was the first generator that had the updated Rein-Seghal model implemented. The new model known as KLN-BS\cite{Nowak:2009se} used the results from the KLN paper \cite{Kuzmin:2003ji} that incorporated the muon mass within the Rein-Sehgal helicity amplitude description and the further improvements from  BS paper \cite{Berger:2007rq} that added the pion pole contributions. The NUANCE and GENIE implementations of \cite{Berger:2007rq} also use the relationships between vector and axial-vector form factors from the Rarita-Swinger formalism and helicity amplitudes approach that were developed by Graczyk and Sobczyk \cite{Graczyk:2007bc}. 

The T2K approach to the analysis uses all available information: the external data from the NA61 to better estimate the neutrino flux, the on-axis detector to monitor the beam stability and event rate and complex near detector (ND280) to constraint systematic uncertainty. The recent improvements in the estimation of the flux using the NA61 data \cite{Abe:2018wpn} allowed to reduce the systematic uncertainty to 8\%-12\%.  The reduction in the flux uncertainty exposed mis-modelling of neutrino interactions in the Monte Carlo generators as flux and cross section predictions are anti-correlated.

The oscillation analysis fitters use new detectors samples to tune the Monte Carlo prediction. T2K uses 14 near detector topological samples, including those with $\pi^0$ and $\pi^+$ in the final states.  The T2K is also the first experiment that attempted to use the CC1$\pi^+$ as the signal. The statistical significance of the addition of this sample is still low but showed that the next generation of experiments could use it\cite{Abe:2017vif}. 

The MiniBooNE experiment had only a single detector and the results \cite{AguilarArevalo:2008rc}  use many the \textit{in situ} measurements, including the pion production channels. The crucial part of the experiment was the constraint for the flux using the HARP data \cite{Catanesi:2007ab}. That allowed for a reduction of the flux uncertainties to a few percent\cite{MiniBooNEFlux}.  The largest backgrounds for the MiniBooNE signal results that come from the pion productions are the asymmetric $\pi^0$ decay from the NC$\pi^0$ channel and the radiative $\Delta \to N \gamma$ production. The fraction of the asymmetric $\pi^0$ decay are well known, and the total $\pi^0$ production was measured by MiniBooNE \cite{AguilarArevalo:2008xs, AguilarArevalo:2009ww}. The upper limit for the NC$\gamma$ production was obtained using the model connecting the $\Delta$ decay to $N\pi^0$ and $N\gamma$. Hopefully, the measurement of this channel will be soon finalized by the MicroBooNE collaboration.

The next few years should deliver many breakthroughs in measurements and analysis of the pion production channels in the sub-GeV neutrino energy region. The SBN (Short-Baseline Neutrino) programme at Fermilab \cite{Antonello:2015lea} will use three detectors to definitely test the existence of the sterile neutrino in the $\Delta m^2 \sim 1$\,ev$^2$ region. The crucial part of the programme is the near detector (SBND), which will collect a huge number of neutrino interactions. In three years the SBND will collect seven millions of $\nu_\mu$ charged current interactions and 50 thousands of  $\nu_\mu$ charged current interactions \cite{Brailsford:2017rxe}. Such statistics will allow for a significant reduction of the systematic uncertainties for the oscillations and cross-section measurements.

\subsection[Impact of RES modeling to the GeV global neutrino oscillation program - Greg Pawloski]{Impact of RES modeling to the GeV global neutrino oscillation program - Gregory Pawloski} \label{sec:2b}
\begin{center}
[\href{https://indico.fnal.gov/event/20793/contribution/5/material/slides/0.pdf}{Presentation}]    
\end{center}
This presentation describes how the modeling of resonance interactions affects the neutrino oscillation program for the NOvA and DUNE experiments.  Both experiments study neutrinos with an energy above 1 GeV, with NOvA looking at a narrow spectrum of neutrino energies that is peaked just below 2 GeV and with DUNE designed to look at neutrino energies in a wide band beam that approximately ranges from sub-GeV energies to 4 GeV.

For both experiments, the primary neutrino interaction process is resonant production.  However, quasi-elastic, 2p2h, and deep (shallow) inelastic interaction modes occur at comparable rates.  The resulting differences in the lepton kinematics and the hadronic activity, among the various production modes, produces differences in the neutrino interaction selection efficiencies and the energy reconstruction.  These effects degrade the sensitivity of the oscillation measurements.  Fortunately, both NOvA and DUNE are capable of directly measuring particles from the hadronic activity which helps reduce the model dependence of the energy reconstruction.  Hence, the largest impact from modeling uncertainties is the effect on selection efficiencies.

\subsection[Impact of RES modeling to the global neutrino-nucleus scattering measurement program - Clarence Wret]{Impact of RES modeling to the global neutrino-nucleus scattering measurement program - Clarence Wret} \label{sec:2c}
\begin{center}
[\href{https://indico.fnal.gov/event/20793/contribution/6/material/slides/0.pdf}{Presentation}]
\end{center}
The global dataset on neutrino-induced single pion production is large and statistically significant enough to probe missing modelling in our generators, such as nuclear effects. This presentation briefly introduces the reader to the single pion production landscape and some previous work on ad-hoc modifications of the process. The talk then focuses on combining historical measurements from bubble chambers on $\text{H}_2$ and $\text{D}_2$ from ANL and BNL\cite{WCR} with modern measurements from MINERvA\cite{Trung-pion, Carrie-pion, Brandon-pion, Altinok-2017} on CH to evaluate areas of tensions in modelling of single pion production\cite{Stowell:2019zsh}. It does so by using NUISANCE\cite{NUISANCE} and a popular version of GENIE\cite{GENIE}, in use by many neutrino cross-section and oscillation experiments. The primary intent of this work is to provide the GeV-scale neutrino scattering and oscillation community with a modified GENIE single pion production model that better describes external data, and expands on the usual GENIE uncertainty parameterisation.

We find that a tuned GENIE model that best describes data from ANL and BNL on $\text{H}_2$ and $\text{D}_2$ does less well than the nominal GENIE model at describing the nuclear target data from MINERvA, implying missing nuclear effects may the culprit. We also explore tensions in describing the various single pion production channels by MINERvA, and find that our tuned GENIE models fail to simultaneously describe CC1/N$\pi^{+}$ and CC1$\pi^0$ measurements. 

A $Q^2$ dependent suppression---possibly originating in unmodelled nuclear effects in GENIE---is developed from this work and previous measurements at MiniBooNE\cite{MINIBOONE-low-q2} and MINOS\cite{MINOS-low-q2} to alleviate the tension in simultaneously describing the nucleon and nuclear target data. We provide GENIE parameters and a $Q^2$ suppression which best describes the whole ensemble of data considered, and tunes for specific pion final states (e.g. tuning only to CC1$\pi^+$ data), including correlation matrices.

We also discuss shortcomings of our method, some inherent to the sparse data releases common to our field at the moment. For instance, experiments failing to provide shape-only covariance matrices and cross-correlations between kinematic distributions of a final state (e.g. the correlations between the one-dimensional CC1$\pi^+$ differential cross-section in $p_\mu$ and the one-dimensional CC1$\pi^+$ differential cross-section in $T_\pi$). We also discuss the need to honestly assess the model dependence of neutrino cross-section measurements in publications to the community, notably the role that efficiency corrections and unfolding procedures may have. Furthermore, it is important to highlight that this work intentionally only slightly improves the pions and nucleons' kinetic energy and angular spectrum. Addressing this will be a topic of future work for the authors.


\section{Models}
\subsection[Berger-Sehgal and Rein-Sehgal models - Konstantin Kuzmin]{Berger-Sehgal and Rein-Sehgal models - Konstantin Kuzmin}\label{sec:3a}
\begin{center}
[\href{https://indico.fnal.gov/event/20793/contribution/7/material/slides/0.pdf}{Presentation}]    
\end{center}
We analyse all available experimental data on the total and differential cross sections for charged-current reactions of $\nu_\mu$ and $\overline{\nu}_\mu$ 1$\pi$ production through baryon resonances in the accelerator experiments at ANL, BNL, FNAL, and CERN.  The data are used to adjust the nucleon axial mass parameter, $M_A$, and fine tune the nonresonance background (NRB) contribution.  The presented result is an update of previous world average value of $M_A = 1.12 \pm 0.03$ GeV~\cite{Kuzmin:2006dh} used in several MC neutrino generators as default or option.
 
In Page 2, we discuss motivation for the study.  To avoid the uncertainties in calculation of NRB and nuclear effects we suggest to extract the phenomenological parameters by fitting the experimental data obtained only with hydrogen and deuterium targets.

  This study is based on extension of the famous Rein-Sehgal model,
  based on the harmonic oscillator quark model in the relativistic formulation
  suggested by Feynman, Kislinger, and Ravndal~\cite{Feynman:1971r}.
  In Pages 3-10, we shortly survey the essential stages in the development
  of the Rein-Sehgal model~\cite{Kuzmin:2006dh,Rein:81,Rein:1987cb,Kuzmin:2003ji,%
  Kuzmin:2004ya,Berger:07,Graczyk:2007bc,Graczyk:2008zz,Hernandez:2007qq,%
  Kabirnezhad:2017,Kabirnezhad:2017jmf,Kabirnezhad:2016nwu,Kabirnezhad:2017dui}
  and propose some modifications to it.

  We illustrate the effects of lepton mass in leptonic current
  (within the so-called KLN model~\cite{Kuzmin:2003ji,Kuzmin:2004ya}),
  pion-pole contribution to the hadronic axial current (KLN-BS model~\cite{Berger:07}),
  and interference of the resonance amplitudes (the latter effect currently neglected in GENIE.)
  We also discuss the importance to keep up-to-date
  the physical parameters of the nucleon and baryon resonances involved into the model.
  The RS model and its extensions use an ambiguous recipe
  for a renormalization of the Breit-Wigner factors.
  We explain the problem and argue why it is better to avoid the renormalization at all.
  The cross sections calculated without renormalization are a few percent lower
  in comparison with these calculated using the standard RS recipe. 
  The standard RS approach represent the non-interfering NRB just by the amplitude of $P_{11}$ 
  production with the Breit-Wigner factor replaced by an adjustable constant $f_{\text{NRB}}$.
  We adjust this constant from a global fit to the experimental data.

  In Pages 11--19, we explain the details of our statistical analysis
  and compare the predicted cross sections with the experimental data.
  The values of $M_A$ and $f_{\text{NRB}}$ are defined from independent fits
  based on non-intersecting data sets: $M_A$ was obtained by fitting the data for the 
  reactions not requiring the NRB in the RS approach, and then, with fixed $M_A$
  the value of $f_{\text{NRB}}$ is defined
  by fitting the remaining of data for reactions sensitive to the NRB contribution.

  We recommend to clarify the RS model and its extensions
  by getting rid of the normalizations of Breit-Wigner distributions.
  The axial mass parameter and adjustable constant for the fine tune of NRB,
  extracted from the global fit for the $\nu_\mu$ and $\overline{\nu}_\mu$ data
  on the total and differential charged-current cross sections
  measured by experiments with hydrogen and deuterium targets,
  are $M_A = 1.176_{-0.067}^{+0.071}$ GeV and $f_{\text{NRB}} = 1.162_{-0.083}^{+0.078}$.
  The values extracted by fitting the neutrino data alone yields the same.
  The much smaller antineutrino data set is not enough
  to obtain reliable results by using it alone.

\subsection[The MK Model - Minoo Kabirnezhad]{The MK Model - Minoo Kabirnezhad}\label{sec:3b}
\begin{center}
[\href{https://indico.fnal.gov/event/20793/contribution/8/material/slides/0.pdf}{Presentation}]    
\end{center}
Neutrino-nucleon interactions that produce a single pion in the final state are of critical importance to accelerator-based neutrino experiments.
These single pion production (SPP) channels make up the largest fraction of the inclusive neutrino-nucleus cross section in the 1-3 GeV range, a region covered by most accelerator-based neutrino beams.\\
The theoretical models \cite{Adler:1968tw,Rein:1987cb,Sato,Hernandez:2007qq} which describe SPP reactions are usually phenomenological in
nature and their predictive power is limited by the precision of SPP neutrino experiments.  That is why the first generation of neutrino SPP models, were relatively crude, since data at the time was very sparse. The long-standing Rein and Sehgal (RS) model \cite{Rein:81} in NEUT and GENIE is a well known example in the resonance region. As we enter the era of precision neutrino physics these models must be updated and improved. 
A big challenge with SPP neutrino interactions is that various resonances with different structures overlap. Therefore, any model like the MK-model will contain a lot of free parameters (to describe the hadronic interaction physics) that should be adjusted to the data. On the other hand, existing neutrino data on “free” nucleons is scarce and it is doubtful that it will be improved. In this situation, the free parameters of the model are fit to exclusive electron (pion) scatter data to constrain the vector (axial) form factors. This approach was used in LP model \cite{LPP} and DCC model \cite{Nakamura:2015rta}, however they didn't evaluate the systematic uncertainty on their parameters. \\
Single pion production (SPP) from a single nucleon occurs when the exchange boson has the requisite four-momentum to excite the target nucleon to a resonance state which promptly decays to produce a final-state pion (resonant interaction), or to create a pion at the interaction vertex (nonresonant interaction). The MK model \cite{Kabirnezhad:2017jmf} provides a full kinematic description of single pion production in the neutrino-nucleon interactions, including resonant and nonresonant interactions in the helicity basis, in order
to study the interference effect. The resonant interactions of MK model are described by RS model. Although the original RS model use a basic dipole form factor the default form factor in MK model is the one introduced by Graczyk and Sobczyk (GS) \cite{Graczyk:2007bc}. The GS form factor is defined by equating the helicity amplitudes (for $\Delta$ resonance) from the RS model with LP model \cite{Lalakulich:2005cs} in order to extract a new form factor for the $\Delta$ resonance. However, There is not an exact solution for those equations since RS model has only one form-factor for all helicity amplitudes while LP model has three form-factors for $\Delta$ helicity amplitudes. The partial solution of these equations did not agree with electron scattering data and the proposed GS form-factors are a combinations of LP form factor that has a better agreements with data. It is important to notice that the GS form-factor is only calculated for $\Delta$ but it is used for all resonances in the original MK model as it was proposed in reference \cite {Graczyk:2007bc}. \\
In the new approach we substitute the vector helicity amplitudes from RS model with LP model with the same form-factors as it is defined in ref. \cite{LPP}. However, the prediction of MK model with LP form-factor overestimate the electron scattering data \cite{pi0,pi+}. This is because the LP model does not include the nonresonant background when the fit was performed in reference \cite{LPP}. Therefore the result form-factor predict larger cross-section when it is applied to the MK model which include the nonresonant background. Therefore, the next step is to fit the vector form-factor of the MK model to the electron scattering data and the  fist step is to parametrise the form factors properly. For this purpose the LP parametrisation for $P_{11}(1232), P_{11}(1440), D_{13}(1520)$ and $S_{11}(1535)$ resonances are used and for the rest of resonances we use dipole form-factors with adjustable coefficients.\\
For nonresonant interaction we keep the Galster form-factor \cite{Gal} but we let the parameters vary in the fitting. We also define an adjustable phase between resonances and nonresonant helicity amplitudes. The fit result of all free parameters to electron scattering data with 1$\sigma$ error are shown in pages 30-36 of the presentation.\\
For this analysis we use exclusive CC single pion production data in electron-proton reaction namely $e p \rightarrow e p \pi^{0}$ and $e p \rightarrow e n \pi^{+}$ channels. 
The goodness-of-fit of the model to data shows that the model is in agreement with the data. Further validation of the constrained model against inclusive electron and pion scattering data sets demonstrates the robustness of the constrained model, and demonstrates is predictive power (page 37).\\
For axial current of the MK model we use dipole form-factors for both resonance and nonresonant interactions and we define a phase between their helicity amplitudes. At low $Q^2$ we can relate the neutrino-nucleon cross-section, which has pure axial contribution at $Q^2=0$, to pion-nucleon cross-section as it is proposed in reference \cite{Nakamura:2015rta}. Therefore we are able to adjust the free parameter in the axial current (i.e. the phases and form-factor's coefficients) to the abundant pion scattering data (page 39).\\

\subsection[Ghent model - Natalie Jachowicz]{Ghent model - Natalie Jachowicz}\label{sec:3c}
\begin{center}
[\href{https://indico.fnal.gov/event/20793/contribution/10/material/slides/0.pdf}{Presentation}]    
\end{center}
The hybrid model for single pion production on the nucleon developed in the Ghent group  aims at  describing the elementary neutrino-induced pion production cross section  over a broad kinematic range. The formalism is the combination of a low-energy model and a high-energy description :\\

\begin{itemize}
\item
  The low-energy model is similar to the description of pion production processes on the nucleon by the Valencia group \cite{Valpi} and is based 
on the combination the contributions of the delta and more massive isospin-1/2 resonances P11(1440), S11(1535), and D13(1520), with the first order background diagrams obtained from the Chiral Perturbation Theory (ChPT) Lagrangian density for the $\pi$N-system.  For the resonances, the s- and u-channel diagrams are included.  The amplitudes are regularized by a Gaussian dipole form factor, controlling the strength of the cross section when moving away from the peak.  This low-energy model is presented in slides 3-5 of the presentation.  There it also becomes clear that for high values of the invariant mass, the non-resonant amplitudes present pathologies due to the fact that only the lowest order diagrams are considered. Taking into account higher order diagrams quickly becomes unfeasible.
\item
This inspired the development of the  model for the higher-energy regime, based on  a Regge approach, which provides the correct dependence of the amplitude at high W, and is presented in slides 6-14.  This approach allows one to take into account a whole class of contributions or 'Regge trajectory' in an efficient and consistent way.
Our approach is based on the procedure for reggeizing the non-resonant background as proposed in Refs. \cite{GLV,KM} for the vector current contributions, which was extended to the axial current. 
\end{itemize}
The low- and high-energy models for the non-resonant contributions are then combined in an effective way as presented on page 15.

Subsequently, the  model is embedded in the nucleus using the relativistic plane wave impulse approximation (RPWIA). The bound nucleons are modeled by relativistic mean field (RMF) wave functions while the final-state pion and nucleon are described by plane waves. This way, the hybrid-RPWIA model is fully relativistic in both operators and  wave functions.  The effect of final state  interactions (FSI) is however not included in this way.  The importance of FSI in different reaction channels is judged by comparing our predictions with NuWro results, as shown on slides 17-25 in a confrontation with experimental data.

Work on implementing distorted final state is in progress, preliminary results are shown on  slide 26-28.

More details on our approach and its application to data were published as \cite{pi1, pi2, pi3}

\subsection[Giessen model - Ulrich Mosel]{Giessen model - Ulrich Mosel}\label{sec:3d}
\begin{center}
[\href{https://indico.fnal.gov/event/20793/contribution/10/material/slides/0.pdf}{Presentation}]
\end{center}
\label{GiBUU_Theory}
Here the theory basis for the description of pion-production and -absorption, implemented in the GiBUU generator, is described.
\subsubsection{Reactions on the nucleon}
Up to invariant masses of about 2 GeV the nucleon exhibits a well-defined resonance structure whereas for higher masses the resonances become increasingly broader and overlap significantly. Following this observation we take all the resonances up to about 2 GeV explicitly into account, while we treat the higher-lying resonance structure by using methods of perturbative QCD (pQCD). Details that go beyond the short following discussions can be found in \cite{LeitnerDiss,Buss:2011mx}.
\paragraph{Resonance region} In this energy range both s-channel (resonance) and t-channel (background) processes contribute. We take both from the MAID2007 analysis of electron-induced pion production on the nucleon. This fixes all the vector couplings and formfactors for all resonances as well as the vector background amplitudes; in this way we avoid using the outdated Rein-Sehgal model. For the axial couplings and formfactors we use PCAC and dipole formfactors, respectively. The one exception is the $\Delta$ resonance where data have allowed to extract more sophisticated formfactors.

The transition currents are explicitly constructed for spin-1/2 and spin-3/2 resonances; Spin-5/2 resonances are approximated by using the 3/2 currents. The explicit forms for these currents are given in Ref.\ \cite{LeitnerDiss,Leitner:2006sp,Leitner:2006ww}. For the very strong $\Delta$ resonance the standard propagator has been used without any ad-hoc changes as in the latest version of the Valencia model.

With this scheme two problems arise: first, the resonance information has to be supplemented with decay probabilities into channels other than $1\pi$. In the resonance region $N^* \rightarrow N + 2\pi$ decays  open at a mass of about 1.5 GeV; the same is true for decays to $N + \rho$ and $N + \eta$. We obtain these partial decay width from the Manley analysis. Second, the axial background terms are undetermined. For these we essentially assume the form of the vector background terms, with a free strength parameter. The latter is fitted to elementary $\nu + N \rightarrow N + \pi$, background terms for other decay branches remain undetermined. This phenomenological determination of the axial background amplitude is required because there exists no theoretical model for resonances beyond the $\Delta$ that could be used as guidance.
\paragraph{pQCD region}
pQCD allows to express the inclusive cross section in terms of the parton distribution functions. The actual final state of an initially excited nucleon in the deep inelastic scattering (DIS) regime, that we assume to start at an invariant mass of about 2 GeV, is determined by the string fragmentation as encoded in the high-energy generator PYTHIA. Since parton distributions are badly known at low $Q^2$ and since DIS events are connected with large momentum transfers we introduce an additional cut-off factor that eliminates low $Q^2$ events as explained in \cite{Gallmeister:2016dnq}.
\subsubsection{Reactions on the nucleus}
\paragraph{Initial reaction}
Neutrino-induced pion production on the nucleus is assumed to proceed as a quasifree process on single nucleons, i.e.\ in 1p1h1$\pi$ reactions; the nucleons are bound and moving in the Fermi sea (for details see \cite{Buss:2011mx}). Further in-medium effects are due to the Pauli-blocking of final states in the resonance decay as well as in the collisional broadening of nucleon resonances. In general, the latter is less essential than the former \cite{Lehr:2001ju}. In the resonance region there are no additional, free parameters, such as formation times; the time-development of pion production is governed only by the resonance width.
\paragraph{Final state interactions}
The final state interactions (FSI) of initially produced hadrons are described by quantum-kinetic transport theory which describes the time-evolution of single particle phase space distributions \cite{Buss:2011mx}. The individual collision rates are determined by experimental hadron-hadron cross sections.
In the very hard DIS regime at high $Q^2$ the wavefunction of the newly formed hadron needs some time to fully develop to its asymptotic state. Here we adopt a model developed by Farrar and Strikman \cite{Farrar:1988me} for the description of color transparency. A quite complete comparison with lepton-induced hadron-production data on nuclear targets, obtained by the HERMES experiment, was performed in Ref.\ \cite{Falter:2004uc,Gallmeister:2007an}. There it was shown that the consistency of the HERMES and EMC data requires a linear rise of the hadron-nucleon cross sections with time after formation.

Both resonance and DIS events contribute to (multi-)pion production. The final state interactions of pions are strong. It is known that 2-body absorption (through $\pi + N \rightarrow N^*,\quad N^* + N \rightarrow NN$) and 3-body absorption contribute to about equal parts. In modeling these processes care has been taken to maintain time-reversal invariance which links pion production to pion absorption \cite{Mosel:2019vhx} in order to avoid unphysical, redundant degrees of freedom and parameters.

The theory has been tested with the help of $\pi^0$ photoproduction data on nuclei \cite{Krusche:2004uw,Buss:2006vh}, with charged pion electro-production data on nuclei \cite{Gallmeister:2007an,Kaskulov:2008ej}, with pion interaction data with nuclei \cite{Engel:1993jh,Buss:2006yk} and with data for pion production with pion and proton beams \cite{Gallmeister:2009ht}. These tests cover incoming-energy regions from a few 100 MeV to 200 GeV.

In all test cases good agreement with experiment is reached. For the neutrino-induced pion production in MiniBooNE the theory gives larger cross sections and different spectral shapes, but it describes the T2K data both on H$_2$O and on CH very well, both in magnitude and in spectral shape.
The MINERvA data are also well reproduced. This comparison will be discussed more in the later section on the GiBUU generator.

\subsection[Valencia Model - Juan Nieves]{Valencia Model - Juan Nieves}\label{sec:3e}
\begin{center}
[\href{https://indico.fnal.gov/event/20793/contribution/11/material/slides/0.pdf}{Presentation}]
\end{center}

The description of the neutrino pion-production and absorption processes in nuclei~\cite{Hernandez:2013jka} begins from a comprehensive  study of the weak pion production off nucleons ~\cite{Hernandez:2007qq,Hernandez:2010bx,Hernandez:2013jka,Alvarez-Ruso:2015eva,Hernandez:2016yfb}.

\subsubsection{Reactions on the nucleon (pion-nucleon invariant masses below 1.4 GeV)}

Besides the dominant $\Delta$ mechanism, we
include also non-resonant contributions required by chiral
symmetry. These chiral background terms were evaluated
using a nonlinear SU(2) chiral Lagrangian, and we
supplemented them with well-known phenomenological form factors introduced in a way that respected both CVC (conservation of the vector current) and PCAC (partial conservation of the axial current)~\cite{Hernandez:2007qq}. All the vector form factors
were determined from pion electroproduction, and for
them, we adopted the values in Ref.\cite{Lalakulich:2005cs}. Axial form factors are mostly unknown. The term proportional to $C_5^A$
is the dominant one, and assuming the pion pole dominance of the pseudo-scalar $C_6^A$ form factor, PCAC relates this latter form-factor to $C_5^A$. We further adopted Adler’s model~\cite{Adler:1968tw} to fix the remaining sub-leading $C_3^A$ and $C_4^A$ contributions. The $q^2-$dependence and normalization of $C_5^A$ is determined from neutrino induced pion production ANL and BNL data, accounting for deuteron effects~\cite{Hernandez:2010bx}.  From the fit, we obtained $C_5^A(0)=1.0\pm 0.1$. This result was 2$\sigma$  below the value of $\sim 1.20$ derived from the off diagonal Goldberger-Treiman relation (GTR). In some fits, 
we unsuccessfully relaxed Adler’s constraints exploring the possibility of extracting some direct information on $C^A_{3,4}(0)$. We showed in \cite{Hernandez:2010bx} that, the available
low-energy data cannot effectively disentangle the different form-factor contributions.

To extend the model to neutrino energies up to
2 GeV, we included in \cite{Hernandez:2013jka} the (small) contribution of the the spin$-3/2$ $D_{13}(1520)$ resonance. Later, the model was partially unitarized by
imposing the Watson theorem in the dominant vector and axial multipoles~\cite{Alvarez-Ruso:2015eva}. The Watson theorem is a
consequence of unitarity and time-reversal invariance. It
implies that, below the two-pion production threshold, the
phase of the electropion or weak pion production amplitude
should be given by the $\pi N \to \pi N$ elastic phase shifts. As a consequence of imposing Watson’s
theorem, the interference between the dominant direct $\Delta$
contribution  and the background
terms changed, and as a result, a larger value ($1.12 \pm  0.11$) for $C_5^A(0)$ was obtained, in better agreement now with the GTR prediction.

Finally, the model has been supplemented in Ref.~\cite{Hernandez:2016yfb} with additional local terms. The aim was to improve the description of the $\nu_\mu n \to \mu^-n\pi^+$ channel, for which most theoretical models give predictions much
below experimental data. This
channel gets a large contribution from the cross $\Delta$ pole (C$\Delta$P) term and then it is sensitive to the spin $1/2$ component of the Rarita-Schwinger covariant $\Delta$ propagator. The description of the $\nu_\mu n \to \mu^-n\pi^+$ channel considerably improves, without affecting the good results we had already obtained in Ref.~\cite{Hernandez:2010bx, Alvarez-Ruso:2015eva}
for the other channels. We found that the C$\Delta$P
amplitude is substantially suppressed and that consistent
$\Delta$ couplings~\cite{Pascalutsa:2000kd} are preferred. Besides, the new  phases needed to satisfy Watson’s theorem are now
much smaller than those obtained in  \cite{Alvarez-Ruso:2015eva}, indicating that the present version without the phases
is closer to satisfying unitarity. Yet,  $C_5^A(0)$ is now larger
by 3.5\% , and it is in remarkable
agreement with the GTR prediction.

As a test of the vector content of the model, we compared its predictions for pion photon and electroproduction in the $\Delta$ region, and we have also confronted these predictions with data. The  model works  quite well and it leads to a fair description of the measured angular and integrated cross sections~\cite{Hernandez:2016yfb,Sobczyk:2018ghy}. 

In the talk, we  also show comparisons of our results  
with the theoretical predictions of the more elaborate,
fully-unitary, dynamical coupled-channel (DCC) model of Ref.~\cite{Nakamura:2015rta}. This latter model involves a total of few
hundred parameters that are fitted to a large
sample ($\ge$ 22300 data points) of $\pi N \to \pi N$ and $\pi^{\pm}p, \gamma p \to
\pi N, \eta N, K\Lambda$ and  $K\Sigma$ measurements. Given the high degree of
complexity of the DCC approach, it is really remarkable
that the bulk of its predictions for electroproduction of
pions in the $\Delta$ region could be reproduced, with a
reasonable accuracy, by our simpler model. The
latter has the advantage that it might be more easily
implemented in the Monte Carlo event generators used
for neutrino oscillation analyses.

We also carried out a careful analysis of the pion angular
dependence of the CC and NC neutrino and antineutrino
pion production reaction off nucleons. We showed
that the possible dependencies on the azimuthal angle
measured in the final pion-nucleon CM system are
$1, \cos\phi_\pi, \cos2\phi_\pi, \sin\phi_\pi$ and $\sin2\phi_\pi$ , and that the two latter ones give rise to parity violation and time-reversal odd correlations in the weak differential cross sections~\cite{Sobczyk:2018ghy}.

\subsubsection{Reactions on the nucleus}

The $1\pi$ production  in both charged and neutral current neutrino-nucleus scattering for neutrino energies below 2 GeV were studied in Ref.~\cite{Hernandez:2013jka}, and the latest results have been presented in this workshop and at the recent NuFacT 2019 by E. Hern\'andez. We use the theoretical model described above, with all theoretical ingredients,  for one pion production at the nucleon level that we
correct for medium effects. The results are incorporated into a cascade program that apart from production also includes the pion final state interaction inside the nucleus~\cite{Hernandez:2013jka}. Besides, in some specific channels
coherent $\pi$ production is also possible and we evaluate its contribution as well,  following the approach of Refs.~\cite{Amaro:2008hd} and \cite{Hernandez:2010jf}. 
In the talk we  compare to theoretical results by the Giessen Group~\cite{Mosel:2017nzk} and  data from the MiniBooNE~\cite{AguilarArevalo:2010bm}, MINER$\nu$A~\cite{Eberly:2014mra} and
T2K~\cite{Abe:2016aoo}  Collaborations. For MiniBooNE and T2K we have integrated the neutrino flux up to 2 GeV neutrino energy, while in the case of MINER$\nu$A, where a cut of 1.4 GeV in the pion-nucleon invariant mass is applied,
we go up to 5 GeV. We get a reasonable reproduction of MiniBooNE data but we overpredict in the case of MINER$\nu$A, though we seem to get right the forward pion data. For the T2K case the agreement is reasonable considering the large errors but compared to central values we lack forward pions in this case.

\subsection[DCC (coupled-channels) models - Toru Sato]{DCC (coupled-channels) models - Toru Sato}\label{sec:3f}
\begin{center}
[\href{https://indico.fnal.gov/event/20793/contribution/12/material/slides/0.pdf}{Presentation}]    
\end{center}
In this talk, we report on the ANL-Osaka dynamical coupled channel model(DCC model)
\cite{Kamano:2013iva,Kamano:2019gtm}.
for the neutrino induced meson production reactions\cite{Sato:2003rq,Nakamura:2015rta}
and analysis of the model in the transition region between resonance production and
deep inelastic scattering.

The DCC model aims to describe neutrino-nucleon reactions in
the resonance region (RES). ($W < 2 \,GeV$ and $Q^2 < 3 \, (GeV/c)^2$, upper limit of $Q^2$ is limitation of
the current DCC model.)
The quark-parton picture is expected to be applicable in the deep inelastic scattering(DIS)($W > 2GeV$ and $Q^2 > 1(GeV/c)^2$) region. Features of the meson production reaction in the RES region are
excitation of many nucleon resonances (for example, 18 $N^*$ and $\Delta$ resonances in the DCC model)
and opening of various meson and baryon states including multi-pion states.
The model of neutrino induced meson production reaction in RES is expected to describe all inelastic channels.
(pages 3-4)

\medskip

Basically two approaches, isobar model and dynamical reaction model, are commonly employed 
to extract the nucleon resonance properties from the pion and photon, electron induced reactions.
ANL-Osaka DCC model is one of the dynamical approaches.
Features of the ANL-Osaka model are (pages 5-10)
\begin{itemize}
\item Effective Hamiltonian consists of 'bare' isobars and meson-baryon potentials in hadron exchange picture.
\item Scattering amplitudes are obtained by solving coupled channel equation in
$\pi N, \eta N, K\Lambda, K\Sigma, \pi\pi N(\pi\Delta,\rho N,\sigma N)$ Fock Space,
which respects  two-body and three-body($\pi\pi N$) unitarity.
\item From the analysis of all available meson production data ,
the parameters of DCC model are determined and the resonance properties are extracted from the amplitudes
of the DCC model.
\item  The model is extended for the electron scattering, the neutron reaction and finaly for the neutrino reaction.
In particular, PCAC is used to determine axial vector coupling constants of resonances and
dipole form is assumed for the nucleon-resonance transition form factors.
\end{itemize}

\medskip

The comparison of the DCC model and the expermental data on total cross sections of
$\pi N$, $\gamma p$, $p(e,e')X$ reactions, $p(\pi, \pi\pi)N$ cross sections
and pion angular distributions of $p(e,e'\pi^0)p$ are shown. The DCC model describes well both
the single pion production reactions and also the total cross sections,
where the two-pion production process is the main inelastic channel.
It is essential to include two-pion productions for inclusive cross sections
around  $W > 1.5 \,GeV$ at large  $Q^2$.(pages 11-15)

The DCC model\cite{Nakamura:2015rta} are compared with the data of neutrino-reactions.
The Total cross sections of single pion production mainly in Delta resonance region
are compared with ANL/BNL data and also Valencia model (HNV model). Both models explans data well.
In addition, a few data on neutrino-induce two-pion production and the shape of the excitation function
($W$ dependence), which shows appreciable contribution of higher resonances except
$\nu p \rightarrow l^- \pi^+ p$. The rough shape of the $Q^2$ dependence of higher resonance is examined
in comparison with data.(pages 16-23)

\medskip

Interesting question is how the DCC model works  close to DIS.
The appropriate experimental data are not available for this purpose.
Here, the DCC model is examined against the structure functions evaluated using PDF.
In the parton picture, the vector current
and the axial vector current contribute with equal strength to the structure function $F_2$, while
it is  non-trivial to realize this feature with models of hadron picture like the DCC.
Our findings of the DCC model are:
\begin{itemize}
\item $F_{2,p}^{em}$ is well described with the DCC model in agreement with $p(e,e')X$ data
and approaches to the $F_{2,p}^{em}$ calculated from PDF in the higher $W$ region.
\item The charged current, proton-neutron averaged $F_2^{CC}$ of the DCC model is about a half of
that of PDF around $W \sim 2 \,GeV, Q^2 \sim 1 \,GeV^2$. The contribution of axial vector current is very small
at higher $Q^2$ region.
\end{itemize}

One of the reasons for the failure would be due to our assumption of dipole form factors for nucleon-resonance
transition form factors.
Since the axial coupling constants at $Q^2 = 0$ are determined assuming PCAC
and the $F_2^{CC}(0)$ the DCC model is consistent with $\pi N \rightarrow X$
and $\pi N \rightarrow \pi N$ cross sections, we have  modified only $Q^2$ dependence of the
transition form factors of 'bare' resonances using those determined from our analysis of the electron scattering.
As a result, we find
\begin{itemize}
\item the 'modified' DCC model predicts $F_2^{CC}$ close to the parton model.
\end{itemize}
This results is encouraging indication on a direction to improve the model at high $Q^2$ and $W$ region.
(pages 24-31)

\medskip

Summary is given in pages 32-33.

\subsection[Duality in neutrino scattering - Jorge Morf\'{i}n]{Duality in neutrino scattering - Jorge Morf\'{i}n}\label{sec:3g}
\begin{center}
[\href{https://indico.fnal.gov/event/20793/contribution/13/material/slides/0.pdf}{Presentation}]    
\end{center}
The Shallow Inelastic Scattering region, particularly the higher-W transition to the Deep-Inelastic Scattering region in $\nu/\nub$ nucleon/nucleus scattering has been scarcely studied theoretically or experimentally. In particular the evolution of low-$Q^2$ non-resonant single pion production to low-$Q^2$ non-resonant multi-pion production and, as $Q^2$ increases, DIS quark-fragmented `pion production needs much more attention.  The lack of knowledge of this region is reflected in the disparity in the current predictions by the community's simulation programs as displayed on slide 2.

There have been multiple studies of the $\Delta$ resonance region (W $\le$ 1.4 GeV), however only restricted studies by the MINERvA 
experiment including somewhat higher W single and  multi-pion production ( W $\le$ 1.8 GeV)~\cite{Stowell:2019zsh} and nothing for the interesting transition to DIS at even higher W.  
 

As shown in this presentation, the application of duality seems to be quite different for e/$\mu$-N interactions  and $\nu/\nub$-N interactions. A brief summary would conclude that:
\begin{itemize}
 \item F$_2^{ep,en}$ - for e/$\mu$-N scattering qualitative and quantitative duality {\emph is observed}    
 \item F$_2^{\nu p,\nu n}$ - for $\nu/\nub$-N scattering duality is roughly observed for the average nucleon [(n+p)/2] but duality is {\emph not observed} for neutrons and protons individually. 
  \item For electroproduction with nuclei it is a different story. The quantitative evaluation of duality in e-A is not as good as with e-N.
   \item For $\nu$-A interactions it is not clear at all how duality works, particularly with nuclei having an excess of neutrons.
 \end{itemize}

The challenge of addressing duality with neutrinos is that in general in the SIS region the resonance structure functions for proton are much larger than for neutrons and in the case of deep-inelastic scattering the opposite is the situation. This does support the observation that if  duality is observed at all with neutrinos it is with the average nucleon $[(n+p)/2]$.

However there is a more fundamental concern regarding the whole concept of testing duality experimentally. Can one really test duality if both the "DIS" and "SIS" regions are not  experimentally accessible at identical kinematics?  For example, the figure on slide 2 represents a neutrino energy typical for the MINERvA experiment and would contribute to the DUNE experiment. There is very limited range of W above the 2 GeV DIS cutoff available for any comparison to the SIS region.  
Furthermore, although there may be limited contributions of higher twist for lower-x and $Q^2$ structure functions, when including inclusive cross sections over all x and $Q^2$ leading twist alone may not be sufficient.  
Thus different extrapolations will give you better or worse agreement between the extrapolated "DIS" part and the measured SIS part.  There is a need for careful consideration of exactly what experimental tests can be made to test duality with neutrino nucleus interactions


This also strongly suggests that rather than only experimental tests of duality  we should encourage a closer examination of just how well the current neutrino simulation event generators, GENIE, NEUT and NuWro obey duality in their treatment of the basic input, $\nu/\nub$ isoscalar nucleon scattering.

\subsection[General Issues with Implementation of Theory Models - Alexis Nikolakopoulos]{General Issues with Implementation of Theory Models - Alexis Nikolakopoulos}\label{sec:3h}
\begin{center}
[\href{https://indico.fnal.gov/event/20793/contribution/20/material/slides/0.pdf}{Presentation}]    
\end{center}
The modeling of neutrino single pion production (SPP)  over the large phase space relevant to neutrino experiments is a problem necessarily complicated by the high dimensionality imposed by the kinematics of the process.
While several approaches have been developed over the past years, many of these models do not readily find their way into neutrino event generators, and when they do it is often with approximations in order to keep the computational cost to a minimum.
 In this talk we look at a number of different models and approximations often made in neutrino event generators. Focusing first on the nucleon, and then on the nuclear degrees of freedom.
\subsubsection{Single pion production on the nucleon}
Kinematically, electroweak production of pions on a free nucleon is a five-dimensional problem. 
In electron scattering experiments the most exclusive cross section has been measured over a large kinematic range.
This wealth of data has been used to determine the couplings to the nucleon resonances and sets stringent constraints on the vector current.
For neutrino interactions no such dataset is available, and one is restricted to using PCAC and pion-pole dominance to constrain the axial couplings at $Q^2=0$.
The $Q^2$ dependence of the dominant Delta (and other resonances) is usually taken as a dipole which is then fitted to the ANL and BNL  data.
These datasets seemed to be incompatible, this discrepancy has been resolved by reanalysis of the data~\cite{Sobczyk2009, Wilkinson2014}.
Recently however it was shown that final-state interactions affect the interpretation of the data even further~\cite{Nakamura2019}.
Also constraints from duality, the idea that the structure functions for hadronic processes should evolve smoothly into the DIS regime for large $W$ and $Q^2$, applied to the Dynamical Coupled Channels (DCC) model of Ref.~\cite{Nakamura:2015rta} seem to suggest that the simple dipole parametrization of the axial form factors is not realistic.
Better results are found when the $Q^2$ dependence of the vector part, extracted from the analysis of electron scattering data, is used~\cite{SatoFF}.

In a recent work~\cite{Sobczyk:2018ghy} the Sato-Lee~\cite{SL}, DCC of Ref.~\cite{Nakamura:2015rta} and the Hernandez-Nieves-Valverde (HNV)~\cite{Hernandez:2010bx} models were rigorously compared, not only for the ANL-BNL datasets, but also in terms of the structure functions for electron and neutrino reactions around the $\Delta$ peak, yielding remarkably similar results. 
In the same work it was argued that neglecting the angular distributions of pions in neutrino event generators could affect the analysis of neutrino experiments.
Indeed, in many neutrino event generators the distribution of outgoing pions is taken to be isotropic in the center-of-momentum system (CMS), or given by some average distribution.
In order to investigate this, we show recent progress made by the NuWro group to implement the angular distributions of pions in neutrino event generators.
The proposed algorithm is based on inverse transform sampling of the structure functions in order to increase the sampling efficiency as compared to the commonly used accept-reject sampling. 
As we do not make any approximations or assumptions based on the components of a specific model, this approach is general and can be used to implement any SPP model in the NuWro MC generator, thereby facilitating a comparison of different models and their effect on the oscillation analysis.

\begin{itemize}
\item Slides 3 - 5: introduction of the kinematics of SPP on the nucleon and definition of the structure functions that determine the angular dependence in CMS.
\item Slides 9 - 12: the structure functions for electroproduction of $\pi^+$ on the proton obtained with the Ghent hybrid model of Ref.~\cite{RGJ2017} are shown and compared to the ones obtained with MAID07~\cite{MAID07} and experimental data.
\item Slides 13 - 15: the structure functions and angular distributions obtained with the HNV, SL, and DCC models as presented in Ref.~\cite{Sobczyk:2018ghy} are compared with the ones obtained with the hybrid model.
\item Slides 16 - 22: an algorithm and possible variations that will be used to implement the angular distributions of pions in NuWro is discussed.
\end{itemize}

\subsubsection{Single pion production on the nucleus}
When SPP takes place on a nucleon bound in the nucleus, 4 additional degrees of freedom are introduced as the final-state nucleus can be left in any state.
Here we will focus on the computation of the hadron current which also gets more involved.
While the distortion of the pion wave function is most certainly important, in this talk we focus only on the nuclear states.

In order to make computations feasible the impulse approximation (IA) is nearly always adopted.
In this case the matrix element is reduced to scattering off independent single-particle states in the nucleus. 
One can further simplify the matrix element by adopting the relativistic plane wave impulse approximation (RPWIA) in which the final state is an on-shell Dirac spinor while the initial state is a bound single-particle state.
In this case the remaining 3-d integral in the current can be evaluated analytically.
The RPWIA matrix element can however not be further factorized, because of the contribution of negative energy components in the bound state wave function~\cite{Caballero1998}.
When these negative energy states are projected out one obtains the plane-wave impulse approximation (PWIA). In this case the matrix element further factorizes in an off-shell free nucleon cross section weighted by the momentum distribution of the nucleus evaluated at the appropriate missing (energy-)momentum.
Results obtained with different energy-momentum distributions (or spectral functions) have recently been compared for neutrino-induced one-nucleon knockout reactions in Ref.~\cite{VanOrden2019}. 
The (R)PWIA approaches clearly falls short in the description of the influence of the nuclear medium on the final state wave function.
To include the effect of final state interactions, the outgoing nucleon can be treated as a scattering state in the same potential used to generate the initial state. 
While this approach provides good results for inclusive electron scattering up to intermediate values of the momentum transfer, the drawback is that the potential is independent on the energy of the outgoing nucleon.
It is well-known that the real potential should get softer as the nucleon's energy increases because more inelastic channels open.
This is clearly seen in $(e,e^\prime p)$ measurements and nucleon nucleus scattering, where phenomenological optical potentials successfully describe the data when taking into account a complex part to account for inelastic reactions which remove the nucleon from the final state.
As a consequence of the increasing complex potential, the real part has to decrease.
To get inclusive cross sections from these potentials different approaches have been used.
In the Relativistic Green's function approach~\cite{Meucci2015} the flux lost in inelastic channels can be recovered.
Another approach is to use only the real part of the phenomenological optical potential such that no flux is lost, this method has seen much use over the past 40 years in describing nuclear reactions.
The approach dubbed the energy-dependent RMF~\cite{RGJ2019a} consist in scaling the RMF scalar and vector potentials with a function dependent on the nucleon's energy, based on an analysis of (e,e') data used to construct the SuSAv2 model. 
The main benefit of this approach is that, by construction, the RMF potential is recovered when the outgoing nucleon's energy is small, thereby respecting the orthogonality of initial and final state wave functions when it is most important.
A comparison of the inclusive scaling function obtained with different energy-dependent potentials and the SuSAv2 approach was presented in Ref.~\cite{RGJ2019b}.

While in recent years we have seen a proliferation of different models for the computation of the hadron current applicable to neutrino pion production on the nucleus (e.g. the ED-RMF of Ref.~\cite{RGJ2019a}, the effective spectral function approach of Ref.~\cite{Rocco2019}, the local Fermi gas with medium corrections of Ref.~\cite{NievesNucleus}, and the GiBUU implementation~\cite{Mosel:2017nzk}), the most common model currently used in MC event generators is the RFG.
While in Ref.~\cite{Praet2009} it was shown that for inclusive observables the RFG provides a description close to the RPWIA, it is unclear how different approximations to the RFG perform. 
Some common approximations are the averaging of the angular dependence of the matrix element (by using e.g. isotropic distributions of pions in CMS), and neglecting the dependence on the energy-momentum of the initial nucleon (by using cross sections computed for stationary nucleons).
In light of the differences found between neutrino-induced SPP datasets and MC generator implementations, which are most obvious at low values of $Q^2$, an in-depth comparison of different nuclear models would be instructive to pin down the origin of this discrepancy.

\begin{itemize}
\item Slides 23 - 29: the kinematics and matrix element are introduced and the assumptions to derive the RPWIA are highlighted.
\item Slides 30 - 35: the PWIA is introduced and different energy-momentum distributions are discussed.
\item Slides 36 - 41: the RDWIA is discussed and the argument for an energy-dependent potential is made. Different approaches of energy dependent treatments are explained.
\item Slides 42 - 46: results for inclusive electron scattering with the ED-RMF approach are shown.
\item Slides 47 - 50: results for neutrino-induced SPP with the (ED-)RMF approach are compared to the RPWIA results to highlight the effect of nucleon FSI. We show results for $Q^2$-distributions with different incoming energies.
\end{itemize}


\section{Implementation in Monte Carlo Simulations}
\subsection[GiBUU - Ulrich Mosel]{GiBUU - Ulrich Mosel}\label{sec:4a}
\begin{center}
\href{https://indico.fnal.gov/event/20793/contribution/18/material/slides/1.pdf}{Presentation}    
\end{center}

The theory of the Giessen model for pion production and absorption (see Sect.\ \ref{GiBUU_Theory}) has been implemented into the quite general generator GiBUU \cite{gibuu} for hadron-, lepton, and photon-induced reactions. GiBUU differs from all the other generators in that it is built on quantum-kinetic transport theory which describes the time-evolution of phase-space distributions of all hadrons involved in the reaction. It also works with momentum-dependent nuclear mean-field potentials so that the initial target nucleus is bound and stable. Furthermore, GiBUU can be used to describe a whole class of nuclear collisions, starting from heavy-ion reactions, as one extreme, and going to neutrino-nucleus reactions, as the other extreme. For all such reactions transport theory and thus the GiBUU generator assumes from the outset that the reaction consists of incoherent elementary processes. Any coherent reactions are thus not described by GiBUU and would have to be added in by hand.

In this generator the physics of $\pi N \Delta$ dynamics in nuclei is used for all the very different reactions; for all such reactions the same physics and code is used. Particular emphasis is placed on maintaining consistency between the various reaction channels, i.e.\ among others using the same groundstate for all channels, and time-reversal invariance which ties pion absorption to pion production in a unique way. In GiBUU this invariance (i.e.~detailed balance) is strictly maintained, as for many other mesons, for the $1\pi$ channel in the resonance region. For the many-pion decay channels in the resonance and DIS regime this invariance is not enforced because it is computationally quite demanding for such channels and, furthermore, the inverse process of simultaneous multi-pion absorption is very rare. 

As discussed in the section on the theory basis of GiBUU, Sect.~\ref{GiBUU_Theory}, the theory and code have been widely tested for all sorts of pion production as well as for pion-induced reactions. For neutrinos the generator has been applied in Refs.\ \cite{Leitner:2008wx,Leitner:2009de,Lalakulich:2010ss}. In particular, in Ref.\ \cite{Leitner:2010kp} we have discussed that QE scattering and pion production are closely entangled and cannot be separated by experiment since the final state for both reactions can be identical because of the FSI. This limits the applicability of theories that give only the quasielastic response. The first extensive set of pion production data for $\nu A$ reactions obtained by MiniBooNE was analyzed in \cite{Lalakulich:2012cj}; it was shown that the data are in general too large compared to theory and that the kinetic energy distribution of pions has an unexpected shape. First predictions for T2K were then published in \cite{Lalakulich:2013iaa}.

Detailed comparisons with the MINERvA data were performed with GiBUU and published in Ref.~\cite{Mosel:2015tja}. Here it could be shown that the MINERvA data could be quite well described. Remaining discrepancies, mainly in the angular distribution, could be attributed to coherent pion production which is is not contained in GiBUU. The very same theory also gave excellent agreement with the T2K pion production data \cite{Mosel:2017nzk}. Since T2K has a flux that is similar to that of MiniBooNE we have concluded that there is no 'pion puzzle' and that the MiniBooNE data for pion production are not correct, both in their absolute cross section and the kinetic energy distribution.

Further support for the treatment of pion production and absorption in GiBU came from studies of so-called $0 \pi$ events in which pions are first created and then reabsorbed again by FSI \cite{Mosel:2017ssx}.

The generator is freely available as a tar-ball from the homepage \texttt{gibuu.hepforge.org}. There also an extensive documentation and helpful explanations can be found. The code runs on any LINUX-based PC and gives many precomputed cross sections. For more detailed analyses, that also take experimental acceptance cuts into account, the code delivers full final state information by giving all the four-momenta and positions of all final state particles at the end of the time-evolution. This final state event file also contains other useful informations for each event, such as, e.g., the type of the initial neutrino-nucleon reaction. The final state file can be obtained in several different output formats, among them in 'root' format and in .txt format.

\subsection[GENIE - Adi Ashkenazi]{GENIE - Adi Ashkenazi}\label{sec:4b}
\begin{center}
[\href{https://indico.fnal.gov/event/20793/contribution/14/material/slides/0.pdf}{Presentation}]    
\end{center}

\subsection[NEUT - Christophe Bronner]{NEUT - Christophe Bronner}\label{sec:4c}
\begin{center}
[\href{https://indico.fnal.gov/event/20793/contribution/15/material/slides/0.pdf}{Presentation}]    
\end{center}

\subsection[NuWro - Kajetan Niewczas]{NuWro - Kajetan Niewczas}\label{sec:4d}
\begin{center}
[\href{https://indico.fnal.gov/event/20793/contribution/16/material/slides/0.pdf}{Presentation}]    
\end{center}


\section{Neutrino, Electron and Pion Scattering Measurements}
\subsection[e-A Scattering Measurements and e4nu - Lawrence Weinstein]{e-A Scattering Measurements and e4nu - Lawrence Weinstein}\label{sec:5a}
\begin{center}
[\href{https://indico.fnal.gov/event/20793/contribution/27/material/slides/0.pdf}{Presentation}]    
\end{center}

\subsection[Other relevant pion scattering measurements - Jacob Calcutt]{Other relevant pion scattering measurements - Jacob Calcutt}\label{sec:5b}
\begin{center}
[\href{https://indico.fnal.gov/event/20793/contribution/25/material/slides/0.pdf}{Presentation}]    
\end{center}

Pion scattering experiments can be used to constrain both Secondary Interaction (SI) and Final State Interaction (FSI) models within neutrino scattering simulations. These two types of interactions affect the ability to reconstruct charged pions produced by neutrino interactions. 

Through FSI, pions produced at the primary vertex of the neutrino interaction can be absorbed within the nucleus. Conversely, pions can be created by a struck nucleon interacting with nuclear matter after the initial neutrino interaction.

Through SI, pions emitted from the nucleus (either at the vertex level or through FSI as described previously) can reinteract with the detector medium. Several types, or topologies, of interactions  can occur, and these affect the ability to reconstruct pions in different ways. 

These processes are accounted for when measuring neutrino oscillations and cross sections. This brings along systematic uncertainties created by a lack of knowledge of their rates or dynamics. Furthermore, it produces difficulties when comparing data to pion production models. A reconstructed pion could have been created through FSI as opposed to a resonance interaction; a pion created by a resonance interaction can be absorbed promptly after existing the nucleus and be missed by reconstruction altogether. Pion scattering experiments can be used to improve model-builders' understanding of these processes.

Historical data exists and has been used by recent experiments such as T2K\cite{T2KFits}, but this talk focused on the recent experiments: Dual Use Experiment at Triumf (DUET), Liquid Argon in a Testbeam (LArIAT), and ProtoDUNE. 

\subsubsection{DUET}

DUET was a $\pi^+$ -- Carbon scattering experiment at TRIUMF. It consisted of a tracking portion (PIA$\nu$O) consiting of scintillating fibers which also served as the target for the experiment, and a downstream scintillator detector known as CEMBALOS\cite{DUET}. Absorption and Charge Exchange interactions were identified within PIA$\nu$O by first identifying an interaction vertex within the fiducial volume of the fibers without a pion-like track exiting the vertex. Then, the Charge Exchange interactions were separated from the Absorption using a selection of events with hits in the CEMBALOS detector\cite{DUET_abs_cex}. 

The results from DUET were used in T2K to constrain their FSI model (specifically, a Cascade model) within their neutrino interaction simulation, NEUT \cite{NEUT}. NEUT's pion scattering routines were used to estimate the (inclusive and exclusive) pion scattering cross sections on various nuclei. Parameters controlling the rates of pion-nucleon level interactions within the Cascade were varied, producing multiple estimations of the cross sections. These were fit to data (including DUET and other measurements), producing a new set of best-fit values and uncertainties for the parameters\cite{T2KFits}.

The DUET results were also used within NO$\nu$A's CC1$\pi^0$ cross section measurement. The signal process for this is a neutrino CC interaction with 1 neutral pion in the final state. A background to this is a final state charged pion promptly interacting and producing a $\pi^0$ through Charge Exchange. The DUET Charge Exchange cross section was used to constrain this background\cite{NOvA}.
\subsubsection{LArIAT}

LArIAT was a hadron scattering experiment which data from a testbeam at FNAL. The detector was a Liquid Argon Time Projection Chamber (LArTPC). The Liquid Argon (LAr) served as both the target and tracking medium. The total cross section for both $\pi^+$ and $\pi^-$ was measured. This measurement is still preliminary, but hints at discrepancies to Geant4's modeling at both high energies and within the resonance region\cite{LArIAT_PiMinus, LArIAT_PiPlus}.

\subsubsection{ProtoDUNE}

ProtoDUNE is another LArTPC residing in a beamline at CERN. ProtoDUNE took beam data in the Fall of 2018 until the Long Shutdown 2 (LS2) began. A second run of beam data is proposed to take place following the end of LS2. 

Work is ongoing to calibrate the ProtoDUNE detector, and to measure the $\pi^+$ cross sections (both total and exclusive) using the beam data.

\subsection[MINERvA capabilities, measurements and plans - Trung Le]{MINERvA capabilities, measurements and plans - Trung Le}\label{sec:5c}
\begin{center}
[\href{https://indico.fnal.gov/event/20793/contribution/21/material/slides/0.pdf}{Presentation}]    
\end{center}

This presentation summarizes MINERvA low-energy resonant pion measurements.

MINERvA detector is a tracking calorimeter located upstream of the MINOS near detector and exposed to the NuMI beam. Muon charge and momentum are measured by the MINOS near detector. Both charged and neutral pion four-vectors can be measured by MINERvA. Charged pion kinetic energy is measured from range while neutral pion energy is measured from calorimetry. The $\pi^+$ is tagged using Michel electron from the decay chain $\pi^+ \rightarrow \mu^+ \rightarrow e^+$. Pions are also separated from protons using a log likelihood ratio score. Two-photon invariant mass is calculated from the two photons to identify neutral pion.

For all four channels of charged-current pion production on hydrocarbon (CH), MINERvA reported differential cross sections as function of observables:~$p_\mu$,$\theta_\mu$,$T_\pi$, and $\theta_\pi$. Differential cross section as function of $Q^2$ and total cross section $\sigma (E_\nu)$ were also reported~\cite{Brandon-pion, Trung-pion, Carrie-pion, Altinok-2017, Trung-2019}. The muon variables are used to probe the primary interaction models since the muon is not sensitive to final-state interactions (FSI). Two charged pion channels, $\pi^+$ for neutrinos and $\pi^-$ for antineutrinos, observed a rate discrepancy of 10-15$\%$ between data and GENIE generator prediction. In general, shape agreement is good for muon observables across all channels.

Pion kinetic energy and angle are used to probe FSI models since the pion is very sensitive to FSI. There are several FSI processes: elastic and inelastic scattering, absorption, and charge exchange. Different channels are sensitive to different combination of these processes. Neutral pion channels are more sensitive to charge exchange. Shape agreement is reasonable between data and GENIE generator.

Total cross sections, $\sigma (E_\nu)$, for neutrinos channels rise to flat-top and stay constant as a function of $E_\nu$ while those for antineutrinos gradually increase with $E_\nu$. These distinct behaviors can be attributed to the interference term $VA$, which contributes constructively and destructively in neutrinos and antineutrinos, respectively.

Differential cross sections as function of $Q^2$ shows very mild suppression at very low $Q^2$ values for $\pi^+$ channel while there is no suppression in the $\pi^-$ channel. Both neutral pion channels show low $Q^2$ suppression, strong evidence from the neutrino $\pi^0$ data, but statistic limited in the antineutrino $\pi^0$ channel. Previous measurements using different targets and neutrino energies reported strong low $Q^2$ suppression \cite{MINOS-low-q2, MINIBOONE-low-q2}  Finally, for the neutrino $\pi^0$ channel MINERvA also measured $p + \pi$ invariant mass, $W_{p + \pi}$. This quantity can be used to study $\Delta (1232)$ dynamics inside the nucleus. Data shows a large shape discrepancy between data and GENIE prediction.

MINERvA have made measurements of all four charged-current pion production channels using the low-energy (3 GeV) datasets. Antineutrino pion production results have limited statistics. MINERvA completed taking a lot more data at medium energy (6 GeV): 4x more neutrino data and 10x more antineutrino data. Results from these high-statistic samples are coming soon.

\subsection[T2K capabilities, measurements and plans - Daniel Cherdack]{T2K capabilities, measurements and plans - Daniel Cherdack}\label{sec:5d}
\begin{center}
[\href{https://indico.fnal.gov/event/20793/contribution/22/material/slides/0.pdf}{Presentation}]
\end{center}

The T2K long-baseline neutrino experiment has a near detector, ND280~\cite{nd280}, located 280 m downstream of the T2K beam target at an off-axis angle of 2.5\textdegree. The off-axis angle provides a narrow band beam peaked at 600 MeV with a suppressed but non-negligible high energy tail. The detector is composed of several subdetector systems, all enclosed in a 0.2 T magnetic field. The Pi-Zero Detector (P0D) is composed of orthogonal scintillator tracking planes interleaved with refillable water layers, and sheets of brass. The two Fine-Grained Detectors (FGDs) are composed of orthogonal scintillator tracking planes (CH), one of which (FGD2) also contains planes filled with water. Downstream of each of the three sub-detector modules are argon gas Time Projection Chambers (TPCs) which can measure track curvature at high resolution, providing sign selection and momentum measurements of tracks that exit the downstream ends of their respective tracking detectors. All of the sub-detector modules are surrounded by electromagnetic calorimeters (ECals). The INGRID detector sits adjacent to ND280 along the beam axis. The cross shaped detector, built of interleaved planes of iron and scintillator strips, is used to monitor the beam direction and intensity. A physics module sits directly upstream of the center module of the cross. Runs were taken with two distinct physics modules; the ``proton module'', composed completely of scintillator strips, and the ``water module'', composed of scintillator strips and water. The three detector schemes allow for statistical subtraction measurements on water with either temporally (P0D and INGRID) or spatially (FGDs) separated data sets.

Four completed measurements will be discussed; one using FGD1 data, one using FGD1 and FGD2 data, one using P0D data, and one using INGRID data. The FGD1 analysis uses a mostly pure hydrocarbon target. The FGD1+FGD2 analysis uses the FGD1 data to constrain the hydrocarbon cross section which is then subtracted from the water plus hydrocarbon cross section measured in FGD2. The P0D analysis uses data from the ``water in'' runs, so the target is composed of hydrocarbon, water, and brass. The INGRID analysis employs a similar technique to the FGD1+FGD2 analysis using the proton and water modules. The two FGD based analyses use NEUT 5.1.4.2 as the input MC (and show comparisons with GENIE 2.6.4), while the P0D and INGRID analyses use NEUT 5.3.3 (and compare with GENIE 2.12.4). Both versions of NEUT use the Rein-Sehgal models (including the original form factors) of pion production, including their models for resonance interaction, non-resonant backgrounds, and coherent pion production. No model for diffractive pion production off hydrogen is included. All of these measurements use data samples collected in Runs I - IV, with an accumulated $1.51\times10^{21}$ POT. A discussion of prospects for future measurements will follow. 

Cross section measurements from the T2K collaboration employ a number of strategies to maximize the impact of and to minimize model dependence of the results. Measurements are focused on flux integrated differential measurements in well reconstructed quantities. The processes measured are based on topological signal definitions. Measurements restrict the ``phase space'' considered. In practice this means redefining the signal as events in  certain regions of kinematic space, both in the analysis variables, and in the quantities that are integrated over. The phase space is limited to regions where the detection and selection efficiencies are well understood, well behaved, and generally flat within any particular analysis bin. Backgrounds are also studied extensively, and ``sideband'' samples are used to measure them with data in the same kinematics regions in which they appear in the signal samples. Fake data studies are used to test the robustness of the measurement procedures to modelled and unmodelled data/MC differences in the signal and background processes. Results are published along with data releases that allow for detailed model comparisons and global fits beyond the scope of the initial publications.

\subsubsection{FGD1}
For the FGD1 (CH target) measurement~\cite{t2k_cc1pi_fgd1} events were selected that contained a forward-going muon and a forward-going charged pion, both of which were required to enter the downstream TPC. Additional proton-like tracks were allowed, but not required. The benefit is a pure sample with minimal muon-pion confusion, and good momentum and angle resolution. The downside is strict kinematical restrictions on momentum and angle ranges for the selected sample. The analysis is performed seven times in different kinematic variables ($d^{2}\sigma/dp_{\mu}d\cos\theta_{\mu}$, $d\sigma/dQ^{2}$, $d\sigma/dp_{\pi}$, $d\sigma/d\theta_{\pi}$, $d\sigma/d\theta_{\pi\mu}$, $d\sigma/d\phi_{\textrm{Adler}}$, $d\sigma/d\theta_{\textrm{Adler}}$). Each of the fits are performed separately, and the number of measured events is not forced to agree between measurements. The agreement between the measured number of events was checked and shown to be well within errors. Fits are performed with a single iteration of the D'Agostini~\cite{DAGOSTINI1995487} unfolding procedure. Given the use of a single iteration of D'Agostini, and the features of the efficiency function there is large potential for bias toward the input signal MC in this analysis.

The double differential measurement in muon kinematics uses the full kinematic phase space of the muon ($0.0<p_{\mu}$ GeV/c, $0.0<\cos\theta_{\mu}<1.0$), since the analysis binning allows for efficiency corrections to be performed in analysis bins. This sample also makes use of a Michelle electron tag for low energy (below tracking threshold) pion identification, extending the pion phase space to all momenta and angles. The pion acceptance over this region is not entirely flat, so this extended phase space comes with additional model dependence, and larger systematic uncertainties related to the efficiency correction. By-eye comparisons indicate that the data agrees fairly well with the NEUT prediction over the full phase space, and sit below slightly the GENIE prediction. 

The six single differential measurements all have a restricted muon phase space of $0.2<p_{\mu}$ GeV/c, $0.2<\cos\theta_{\mu}<1.0$. The pion phase space is similarly limited to the $0.2<p_{\mu}$ GeV/c, $0.2<\cos\theta_{\mu}<1.0$ range, except when the variable in question is measured. Again, by-eye comparisons if the data with NEUT show good agreement, other than the well documented discrepancy at low $Q^{2}$ ($Q^{2}<0.2$ (GeV/c)$^{2}$). The data also falls below the NEUT MC at lower pion momenta and small angles, although mostly within errors. The over-prediction of GENIE persists as well, with the excess events concentrated in the region of $\theta_{\pi}>0.8$ rad, $\theta_{\pi\mu}>0.7$ rad, and roughly flat over the the $Q^{2}$, $p_{\mu}$, and Adler angle distributions.

\subsubsection{FGD1+FGD2}
The FGD1+FGD2 (water target) measurement~\cite{t2k_cc1pi_fgd1fgd2} employs statistical subtraction techniques to measure the pion production cross section on water (the target used in Super-Kamiokande, the T2K far detector). The analysis selected events with forward-going muons and pions above the tracking threshold for the FGDs leading to phase space restrictions of $\cos\theta_{\mu/\pi}>0.3$ and $p_{\mu/\pi}>200$~MeV/c. Independent fits in muon and pion $p-\theta$ were performed. Similar to the FGD1 measurement, the non-flat efficiency function and the single D'Agostini iteration unfolding leave potential for bais toward the input signal MC. 

The results in muon kinematics agrees well with the NEUT prediction, sitting just under the predicted spectra, with the error bars covering the difference. The error bars, however, do not cover the difference between the result and the GENIE prediction, which lies well above the NEUT prediction across the full $p_{\mu}$ and $\theta_{\mu}$ ranges explored. The deficit between the data and the NEUT prediction is more concentrated in the pion kinematic space, with the data falling below the MC in the $300<p_{\pi}<700$~MeV/c range and at angles of $\cos\theta_{\pi}>0.94$. This deficit is consistent with the one observed in the FGD1 only analysis.

\subsubsection{P0D}
The P0D (CH, water and brass target) analysis~\cite{t2k_cc1pi_p0d} selects events with two forward-going MIP-like tracks. Tracks must either exit the P0D and enter the TPC, or be contained in the P0D. The analysis is done in bins of $p_{\mu}-\theta_{\mu}$ for events with forward-going pions with momentum above 250~MeV/c. No unfolding is used and the results are presented in the reconstructed kinematic space. The muon efficiency is fairly flat across each analysis bin, and the pion efficiency is flat above the 250 MeV/c threshold and over all forward-going angles. The well-behaved efficiencies and lack of unfolding greatly reduce the potential for signal model dependent bias in the result. However, given the complex target and the need to forward fold models, comparison beyond those provided in the analysis will be difficult. Analysis bins run from $0.14<p_{\mu}<5.0$ GeV/c for all forward-going muon angles. The result is present in 1-D projections due to limited statistics. 

The data was compared with the nominal NEUT MC, as well as versions of NEUT where the resonant (and non-resonant background) production model was changed to the MK model~\cite{Kabirnezhad:2017jmf}, a version of NEUT where the coherent production model was changed to the Berger Sehgal model~\cite{Berger:2007rq}, and a version of NEUT where both changes had been made. Each change reduced the NEUT prediction somewhat. The MK model predicts a softer pion momentum spectrum which shifted events below the 250 MeV/c threshold, while the Berger-Sehgal model greatly reduces the number of events producing pions with a kinetic energy below 1 GeV, which dominate the sample. The data sit well below the nominal NEUT prediction and agree, within errors, with the NEUT version where both models have been changed.

\subsubsection{INGRID}
Due to the physical dimensions of the detector, the design of the scintillator bars, and the lack of a magnetic field, the INGRID (CH and water target) analysis~\cite{t2k_cc1pi_ingrid} uses an even more restricted phase space than the other analyses. Namely tracks must have a momentum above 400 MeV/c, and an angle less than 50\textdegree. The lack of a downstream TPC and magnetic field also limit the ability to measure the momentum of tracks that exit the detector. This means that all tracks above $1.0$ GeV/c are grouped into a single bin. Given the higher energy flux that INGRID sees with its on-axis position, this bin contains a significant number of the observed events. The analysis was performed in the $p_{\mu}-\theta_{\mu}$ space, integrating over all pion tracks in the allowed phase space. Measurements were made using the proton module data on its own, and on water using the water and proton module data and the statistical subtraction technique. A method for determining the number of D'Agostini unfolding iterations based on data was developed for this analysis, reducing the model dependence of the result. The efficiency is fairly flat over the phase space used in the analysis which also reduced model dependency.

The measurement using only the proton (CH) module are mostly consistent with NEUT at higher angles and energies, but sit well beneath the NEUT prediction at low energies for forward going muon tracks. As with the previous measurements, the GENIE prediction is above the 1 $\sigma$ error band across most of the measured kinematic regions. The features of the water measurements are quite different, in that they agree with NEUT in all regions except for the highest angle bin. The errors in this analysis are notably larger due to the statistical subtraction.

\subsubsection{Future Prospects}

T2K has several in progress measurements. Each of them will make use of increased data running (including $\bar{\nu}$ running), improved reconstruction, updated models and model uncertainties, and new unfolding techniques that remove model dependence. Newer analysis will also include combined fits that better, and more consistently, measure and constrain background processes. The reconstruction improvements include a 3-track ($1\mu1\pi1p$) selection that allows for transverse variable analyses. Improvements also include a momentum reconstruction of pions from Michele electrons that is being developed, and which has the potential to push down the charged pion thresholds to near zero. In addition, improved track reconstruction for high angle tracks in the FGDs, which has already been applied to $0\pi$ measurements, are currently being implemented in single pion production measurements. Work on multiple $\bar{\nu}$ production measurements is underway, looking at CH and water targets, as well as a coherent focused analysis. The new unfolding technique does not use D'Agostini, and allows for direct control over the regularization strength to ensure minimal model dependence. Greater attention has also been payed to efficiency functions at all levels of analysis development which will reduce model dependent bias in all future results.

\subsection[NOvA capabilities, measurements and plans - Leonidas Aliaga]{NOvA capabilities, measurements and plans - Leonidas Aliaga}\label{sec:5e}
\begin{center}
[\href{https://indico.fnal.gov/event/20793/contribution/23/material/slides/0.pdf}{Presentation}]
\end{center}

The high neutrino beam and its energy peaked around 2 GeV at the NOvA near detector provide an excellent opportunity to measure cross sections in the resonance region. The beam is mostly composed by pure muon neutrinos, $\sim$ 96\%,  with a $\sim$ 1\% electron neutrino component. This high statistics offer to most of our analyses of the interaction channels, including the electron neutrino interactions, rich data sets for precise and only systematic limited measurements, that is complementary to the current efforts by other experiments.
  
The NOvA ND is located 100 m underground at Fermilab, 14 mrad off-axis with respect to the average pion decay vertices and about 1 Km from the NuMI target. It is comprised of PVC cells filled with liquid scintillator, alternating planes of orthogonal views, 193 ton of fully active region and 97 ton downstream muon catcher to range out the muon energy.

Most of our analyses measure flux-integrated differential cross sections with a minimal model dependency. The selection cuts are optimized by looking at the minimum value of the approximate expected fractional uncertainty on the cross section coming from the detector response and the interaction model systematics. The simulation is tuned to account for mis-modeling of the hadron production in the beamline simulation to predict the flux and to account for the neutrino-nucleus interaction and nuclear effects mis-modeling \cite{novatune}. The systematic uncertainties are calculated and propagated by three procedures: creating an statistical ensemble of individual randomly generated universe by  shifting any parameter source (flux and interaction model systematics), shifting the calibration parameters and redoing the whole reconstruction chain (detector response systematics) or by shifting variables at the event level (such as muon energy scale, muon angle systematics, etc). 

Signals are defined by the topology of the events with the interaction vertices in a fiducial volume. The events has to be contained in a sub-volume inside the fully active region to avoid hadronic shower leaking. Only muons are allowed to travel to the muon catcher when selecting the muon neutrino events. Backgrounds are treated differently depending of the analysis: using either sidebands, template fits or the simulation values as it is described in the presentation. The event candidates are unfolded to move from the reconstruction to true space, they are efficiency corrected and normalized by the number of target nucleons and the flux. 

This talk presents the status of the cross-section program at the NOvA near detector (ND) and it is particularly focused on detector capabilities and strategies we are following in our analyses to measure cross sections. It is structured showing the analyses lined up with the development and understanding of new tools that improve them: measurements of neutral pion production, progress on measuring inclusive cross sections and current efforts towards other semi-inclusive channel analyses. 

Slides 17-26 show the results on a couple of neutral pion  measurements. First, the neutral-current coherent neutral-pion production (NC Coh $\pi^{0}$) \cite{cohpi0}. This channel measured the flux-averaged total cross section, showing a good agreement with GENIE’s Rein-Sehgal model prediction. The second result is the muon-neutrino semi-inclusive neutral-pion production. This channel measured the flux-averaged cross section as a function of muon and neutral pion kinematics (angle with respect to the beam and momentum), $Q^{2}$ and W. 

Slides 27-43 show the muon- and electron- neutrino inclusive analyses \cite{nuint18_aliaga, nuint18_judah}. First, the muon neutrino analysis measures the double differential cross section as a function of muon kinematics ($cos\theta_{\mu}$ with respect of the beam angle and the muon kinetic energy, $T_{\mu}$). This analysis is done completely in three dimensions: the muon kinematic variables  and the ``available energy`` (the total energy of all observable final state hadrons) \cite{neutrino2020_linda}. The electron neutrino analysis measures, for the first time in the neutrino community, the double-differential cross sections as function of the electron kinematics ($(E_{e}$) and the $cos\theta_{e})$ with respect of the beam angle. 

Slides 44-54 discuss the progress and future directions on some semi-inclusive channels, particularly three ongoing analyses in NOvA: charged-current charged-pions, muon-antineutrino neutral-pions and muon-neutrino with not pions in the final state. These three are shown as examples between many other analyses currently in progress.

\subsection[MicroBooNE capabilities, measurements and plans - Kirsty Duffy]{MicroBooNE capabilities, measurements and plans - Kirsty Duffy}\label{sec:5f}
\begin{center}
[\href{https://indico.fnal.gov/event/20793/contribution/24/material/slides/0.pdf}{Presentation}]    
\end{center}

Resonant interactions form significant parts of the signal and backgrounds to MicroBooNE's signature Low-Energy Excess (LEE) analysis, in ways that are also strongly dependent on Final State Interactions (FSI). The LEE signal topology is CC0$\pi$, of which a large component ($\mathcal{O}(15\%)$ but model dependent) is resonant pion production in which the pion is absorbed in the nucleus. Backgrounds come from neutral pion production (in particular, $\pi^0$ production forms 80\% of backgrounds to a photon-like LEE search), and charged pion production in which the charged pion is misidentified as a muon.

Extra motivation to study resonant interactions in MicroBooNE comes from DUNE: because of the energy regime in which DUNE sits, CC pion production will be an important oscillation signal channel for DUNE, and many of the backgrounds to DUNE oscillation searches will be similar to those in the MicroBooNE LEE. MicroBooNE's high-statistics measurements of neutrino interactions on argon and the particle identification techniques in a liquid argon TPC developed on MicroBooNE will be an important input to future DUNE analyses.

This talk presents MicroBooNE's existing and planned measurements in three categories: measurements that aren't specifically in the resonance region (a CC-inclusive cross section measurement and prospects to measure exclusive final states with protons), measurements of neutral pions, and prospects for measurements with charged pions.

The CC-inclusive measurement published in~\cite{PhysRevLett.123.131801} is presented in slides 18--26, and represents the largest-ever published sample of neutrino interactions on argon. The results are compared to a number of generators: GENIE v2+MEC, GENIE v3, GiBUU, and NuWro. We see tension between the data and all generators, although the tension is reduced for GENIE v3, GiBUU, and NuWro compared to GENIE v2+MEC (the generator used to perform this analysis). Further investigation shows that the tension is driven by the high-momentum, forward-going bins, which are populated largely by QE and MEC processes: the data lie below the predictions, with very small uncertainties. Studies using the GENIE generator show the sensitivity of this inclusive selection to the underlying interaction models. The main reasons behind improved agreement in GENIE v3 compared to GENIE v2+MEC are due to moving from an RFG nuclear model and Llewelyn-Smith CCQE model to LFG and Nieves CCQE, and the inclusion of RPA effects. Both of these changes give suppression in the high-momentum, forward-going region, although we see evidence that more suppression is needed.

Slides 27--30 discuss upcoming (but currently not public) measurements of more exclusive final states including protons, demonstrating a 47~MeV kinetic energy threshold for detecting and identifying protons (currently limited by reconstruction, and expected to improve further in future analyses). Initial selections for CC0$\pi$2p and CC0$\pi$Np (N$\geq$1) are described in~\cite{MICROBOONE-NOTE-1056-PUB}. In the further future, MicroBooNE plans to use these selections to make detailed studies of nuclear effects using Single Transverse Variables. In a large nucleus such as argon, all of these CC0$\pi$ measurements will need to be interpreted using good resonance and FSI models.

Photons and electrons both produce showers in a liquid argon TPC, and NC$\pi^0$ production will be a significant background to any analysis that looks for electron neutrinos. MicroBooNE's published measurement of CC$\pi^0$ production is presented in slides 32--38, along with planned improvements. The published result~\cite{PhysRevD.99.091102} (the first ever CC$\pi^0$ cross section measurement on argon) is a single-bin total cross-section measurement, and shows agreement with GENIE Rein-Sehgal and Berger-Sehgal models, and with NuWro, within 1.2$\sigma$. Comparisons between the same models and measurements on deuterium and carbon imply that the current nuclear scalings are sufficient, within the uncertainties of these measurements. MicroBooNE plans to update this measurement in the near future, to produce a differential cross section as a function of $\pi^0$ momentum, $\pi^0$ angle, muon momentum, and muon angle. Slides 39--42 give a brief overview of existing ArgoNeuT measurements of the NC$\pi^0$ channel on argon~\cite{PhysRevD.96.012006} and the selection currently being developed for use as a sideband in MicroBooNE's LEE analysis.

Finally, slides 44--62 discuss the difficulties with identifying charged pions (mostly $\pi^+$) in liquid argon detectors, largely based on the MicroBooNE experience (although ArgoNeuT measurements of CC1$\pi^{\pm}$ production~\cite{PhysRevD.98.052002} are also shown). Finding a way to identify and characterize $\pi^+$ is an ongoing effort on MicroBooNE. Charged pions cannot be distinguished from muons using charge deposition; the only sure identification of a charged pion is through hadronic interactions in the argon, but this brings its own challenges in terms of reconstruction and measuring the particle's energy.

To summarize, MicroBooNE needs to understand resonant interactions (and FSI) for the success of our own physics program, especially $\pi^0$ backgrounds and the RES+FSI part of the CC0$\pi$ signal. Our measurements can help, as well as provide input to future experiments. We have a number of relevant measurements already published, in progress, or planned, and we appreciate input from the community on the most interesting measurements to make. In order to make precise measurements, MicroBooNE needs good predictions of pion-Argon cross sections (both for understanding FSI and secondary interactions in the detector) and good models for neutrino interactions with realistic and believable uncertainties.

\end{document}